\documentclass[nofootinbib,prd,twocolumn,showpacs,showkeys,preprintnumbers]{revtex4-1}
\usepackage{hyperref,amssymb,amsmath,mathrsfs,bm,graphicx}
\begin{document}
\title {Non--static fluid spheres admitting a conformal Killing vector: Exact solutions}
\author{L. Herrera}
\email{lherrera@usal.es}
\affiliation{Departament de  F\'{\i}sica, Universitat Illes Balears, E-07122 Palma de Mallorca, Spain and Instituto Universitario de F\'isica
Fundamental y Matem\'aticas, Universidad de Salamanca, Salamanca 37007, Spain.}
\author{A. Di Prisco}
\email{adiprisco56@gmail.com}
\affiliation{Escuela de F\'{\i}sica, Facultad de Ciencias,
Universidad Central de Venezuela, Caracas 1050, Venezuela.}
\author{J. Ospino}\affiliation{Departamento de Matem\'atica Aplicada and Instituto Universitario de F\'isica
Fundamental y Matem\'aticas, Universidad de Salamanca, Salamanca 37007, Spain.}
\email{j.ospino@usal.es}
\date{\today}
\begin{abstract}
We carry on a general study on non--static spherically  symmetric  fluids admitting a  conformal Killing vector (CKV). Several families of exact analytical solutions are found for different choices of the CKV,  in  both,  the dissipative and the adiabatic regime. To specify the solutions, besides the fulfillment of the junction conditions on the boundary of the fluid distribution,   different conditions are imposed,  such as  vanishing complexity factor and quasi--homologous  evolution. A detailed analysis  of the obtained solutions, its prospective applications  to astrophysical scenarios, as well as alternative approaches to obtain new solutions, are discussed.

\end{abstract}
\date{\today}
\pacs{04.40.-b, 04.40.Nr, 04.40.Dg}
\keywords{Relativistic Fluids, spherical sources, self--similarity, interior solutions.}
\maketitle

\section{Introduction}
The purpose of this work is twofold. On the one hand we want to delve deeper into the physical consequences derived from the assumption that a given space--time admits a CKV. This interest  in its turn is motivated by the relevance of  such kind of symmetry in hydrodynamics.

Indeed, in general relativity, self--similar solutions are related to the existence of a homothetic Killing vector field (HKV),  a generalization of which is  a conformal Killing vector field (CKV). The physical interest of systems admitting a CKV is then suggested by the important role played by self-similarity in classical hydrodynamics.

Thus, in Newtonian hydrodynamics, self--similar solutions are those described by means of physical quantities which are functions depending on dimensionless variables $x/l(t)$, where $x$ and $t$ are independent space and time variables and $l$ is a time dependent scale. Therefore  the spatial distribution of the characteristics of motion remains similar to itself at all times \cite{1}. In other words, self--similarity is to be expected whenever the system under consideration possesses no characteristic length scale.

The above comments suggest that  self--similarity plays an important role in the study of systems close to the critical point, where the correlation length becomes infinite, in which case different phases of the fluid (e.g. liquid--vapor) may coexist, the phase boundaries vanish and density fluctuations occur at all length scales. This process may be observed in  the critical opalescence.

Besides, examples of self--similar fluids may be found in the study of strong explosions and thermal waves \cite{2, 3, 4, 5}.

Motivated by the above arguments many authors, since the pioneering work by Cahill and Taub \cite{6}, have focused their interest in  the problem of self--similarity in self--gravitating systems. Some of them are restricted to general relativity, with especial emphasis  on the ensuing consequences from the existence of HKV  or CKV, and possible solutions to the Einstein equations (see for example \cite{7, 8co, 9co, 10co, 11co, 12co,13co, 14, 15, 16, 17, 18, 19, 20, 21, 22, 23, 24, 25, 26, 27, 28, 29, 30, 31, 32, 33, 33b, 34, 35, 36, 37, 38, 39, 40, 40bis, 41, 42, 43, 44, sherif, matondo} and references therein). Also, a great deal of work has been done in the context of other theories of gravitation (see for example \cite{rej, RS, MH2, HM2, SIF, SHS, Bhar2, TD, DRGR2, OS, ZSRA, DRGR, SNN} and references therein). Finally, it is worth mentioning the interest of this kind of symmetry related to the modeling of wormholes (see \cite{BHL1, BHL, RRKKK, K, SIF2, kar, sahoo} and references therein).

On the other hand, the problem of general relativistic gravitational collapse has attracted the attention of researchers since the seminal paper by Oppenheimer and Snyder. The origin of  such interest resides in  the fact that the gravitational collapse of massive stars represents one of the few observable phenomena where general relativity is expected to play a relevant role.  To tackle such  a problem there are two different approaches: Numerical methods or analytical exact solutions to Einstein equations.
Numerical methods enable researchers to investigate systems that are extremely difficult to handle analytically.   However,  purely numerical solutions usually hinder the investigation of  general, qualitative, aspects of the process. On the other hand, analytical solutions although  are generally found either for too simplistic equations of state and/or under additional heuristic assumptions whose justification is usually uncertain, are more suitable for a general discussion and seem  to be useful to study non--static models which are relatively simple to analyze but still contain some of the essential features of a realistic situation.

In this manuscript we endeavor to find exact, analytical, non--static solutions admitting a CKV, including dissipative processes. The source will be represented by an anisotropic fluid dissipating energy in the diffusion approximation. In order to find the solutions we shall specialize the CKV to be either space--like  (orthogonal to the four--velocity), or time--like (parallel to the four--velocity).  In each case we shall consider separately the dissipative and non--dissipative regime. Also, in order to specify the models, we  will assume specific restrictions on the mode of the evolution, (e.g. the quasi--homologous condition),  and  on the complexity factor, among other conditions. A fundamental role in finding our models is played by the equations ensuing from the junction conditions on the boundary of the fluid distribution, whose integration provides one of the functions defining the metric tensor.

Several families of solutions are found and discussed in detail.  A summary of the obtained results  and a discussion on the physical relevance of these solutions are presented in last section. Finally several appendices are  included containing useful formulae.

\section{The metric, the source and relevant equations and variables}
In what follows we shall briefly summarize the definitions and main equations required for describing spherically symmetric dissipative fluids. We shall heavily rely on \cite{epjc}, therefore we shall omit many steps in the calculations, details of which the reader may  find in \cite{epjc}.

We consider a spherically symmetric distribution  of collapsing
fluid,   bounded by a spherical surface $\Sigma$. The fluid is
assumed to be locally anisotropic (principal stresses unequal) and undergoing dissipation in the
form of heat flow (diffusion approximation).

The justification to consider anisotropic fluids is provided by the fact that pressure anisotropy is produced by many different physical phenomena of the kind expected in  gravitational collapse scenario (see \cite{report} and references therein). Furthermore  we expect that the final stages of stellar evolution should be accompanied by intense dissipative processes, which, as shown in \cite{ps}, should produce pressure anisotropy.

Choosing comoving coordinates, the general
interior metric can be written as
\begin{equation}
ds^2=-A^2dt^2+B^2dr^2+R^2(d\theta^2+\sin^2\theta d\phi^2),
\label{1}
\end{equation}
where $A$, $B$ and $R$ are functions of $t$ and $r$ and are assumed
positive. We number the coordinates $x^0=t$, $x^1=r$, $x^2=\theta$
and $x^3=\phi$. Observe that $A$ and $B$ are dimensionless, whereas $R$ has the same dimension as $r$.

The  energy momentum tensor  in the canonical form, reads
\begin{equation}
T_{\alpha \beta} = {\mu} V_\alpha V_\beta + P h_{\alpha \beta} + \Pi_{\alpha \beta} +
q \left(V_\alpha K_\beta + K_\alpha V_\beta\right), \label{Tab}
\end{equation}
with
$$ P=\frac{P_{r}+2P_{\bot}}{3}, \qquad h_{\alpha \beta}=g_{\alpha \beta}+V_\alpha V_\beta,$$

$$\Pi_{\alpha \beta}=\Pi\left(K_\alpha K_\beta - \frac{1}{3} h_{\alpha \beta}\right), \qquad \Pi=P_{r}-P_{\bot}$$

where $\mu$ is the energy density, $P_r$ the radial pressure,
$P_{\perp}$ the tangential pressure, $q^{\alpha}=qK^\alpha$ the heat flux, $V^{\alpha}$ the four--velocity of the fluid,
and $K^{\alpha}$ a unit four--vector along the radial direction.
Since we are considering comoving observers, we have
\begin{eqnarray}
V^{\alpha}&=&A^{-1}\delta_0^{\alpha}, \;\;
q^{\alpha}=qK^{\alpha}, \;\;
K^{\alpha}=B^{-1}\delta^{\alpha}_1.
\end{eqnarray}

 These quantities
satisfy
\begin{eqnarray}
V^{\alpha}V_{\alpha}=-1, \;\; V^{\alpha}q_{\alpha}=0, \;\; K^{\alpha}K_{\alpha}=1,\;\;
K^{\alpha}V_{\alpha}=0.
\end{eqnarray}

It is worth noticing that we do not explicitly add bulk or shear viscosity to the system because they
can be trivially absorbed into the radial and tangential pressures, $P_r$ and
$P_{\perp}$, of the collapsing fluid (in $\Pi$). Also we do not explicitly  introduce  dissipation in the free streaming approximation since it can be absorbed in $\mu, P_r$ and $q$. 

The Einstein equations for (\ref{1}) and (\ref{Tab}), are explicitly written  in Appendix A.

The acceleration $a_{\alpha}$ and the expansion $\Theta$ of the fluid are
given by
\begin{equation}
a_{\alpha}=V_{\alpha ;\beta}V^{\beta}, \;\;
\Theta={V^{\alpha}}_{;\alpha}, \label{4b}
\end{equation}
and its  shear $\sigma_{\alpha\beta}$ by
\begin{equation}
\sigma_{\alpha\beta}=V_{(\alpha
;\beta)}+a_{(\alpha}V_{\beta)}-\frac{1}{3}\Theta h_{\alpha\beta}.
\label{4a}
\end{equation}

From  (\ref{4b}) we have for the  four--acceleration and its scalar $a$,
\begin{equation}
a_\alpha=a K_\alpha, \;\; a=\frac{A^{\prime}}{AB}, \label{5c}
\end{equation}
and for the expansion
\begin{equation}
\Theta=\frac{1}{A}\left(\frac{\dot{B}}{B}+2\frac{\dot{R}}{R}\right),
\label{5c1}
\end{equation}
where the  prime stands for $r$
differentiation and the dot stands for differentiation with respect to $t$.

We obtain
for the shear (\ref{4a}) its non zero components
\begin{equation}
\sigma_{11}=\frac{2}{3}B^2\sigma, \;\;
\sigma_{22}=\frac{\sigma_{33}}{\sin^2\theta}=-\frac{1}{3}R^2\sigma,
 \label{5a}
\end{equation}
and its scalar
\begin{equation}
\sigma^{\alpha\beta}\sigma_{\alpha\beta}=\frac{2}{3}\sigma^2,
\label{5b}
\end{equation}
where
\begin{equation}
\sigma=\frac{1}{A}\left(\frac{\dot{B}}{B}-\frac{\dot{R}}{R}\right).\label{5b1}
\end{equation}

Next, the mass function $m(t,r)$  reads
\begin{equation}
m=\frac{R^3}{2}{R_{23}}^{23}
=\frac{R}{2}\left[\left(\frac{\dot R}{A}\right)^2-\left(\frac{R^{\prime}}{B}\right)^2+1\right].
 \label{17masa}
\end{equation}

Introducing the proper time derivative $D_T$
given by
\begin{equation}
D_T=\frac{1}{A}\frac{\partial}{\partial t}, \label{16}
\end{equation}
we can define the velocity $U$ of the collapsing
fluid  as the variation of the areal radius with respect to proper time, i.e.
\begin{equation}
U=D_TR \;\; \mbox{(negative in the case of collapse)}, \label{19}
\end{equation}
where $R$ defines the areal radius of a spherical surface inside the fluid distribution (as
measured from its area).

Then (\ref{17masa}) can be rewritten as
\begin{equation}
E \equiv \frac{R^{\prime}}{B}=\left(1+U^2-\frac{2m}{R}\right)^{1/2}.
\label{20x}
\end{equation}
Using (\ref{20x}) we can express (\ref{17a}) as
\begin{equation}
4\pi q=E\left[\frac{1}{3}D_R(\Theta-\sigma)
-\frac{\sigma}{R}\right],\label{21a}
\end{equation}
where   $D_R$ denotes the proper radial derivative,
\begin{equation}
D_R=\frac{1}{R^{\prime}}\frac{\partial}{\partial r}.\label{23a}
\end{equation}
Using (\ref{12})-(\ref{14}) with (\ref{16}) and (\ref{23a}) we obtain from
(\ref{17masa})
\begin{eqnarray}
D_Tm=-4\pi\left(P_rU+ qE\right)R^2,
\label{22Dt}
\end{eqnarray}
and
\begin{eqnarray}
D_Rm=4\pi\left(\mu+q\frac{U}{E}\right)R^2,
\label{27Dr}
\end{eqnarray}
which implies
\begin{equation}
m=4\pi\int^{r}_{0}\left( \mu +q\frac{U}{E}\right)R^2R^\prime dr, \label{27intcopy}
\end{equation}
satisfying the regular condition  $m(t,0)=0$.

Integrating (\ref{27intcopy}) we find
\begin{equation}
\frac{3m}{R^3} = 4\pi {\mu} - \frac{4\pi}{R^3} \int^r_0{R^3\left(D_R{ \mu}-3 q \frac{U}{RE}\right) R^\prime dr}.
\label{3m/R3}
\end{equation}
\\

\subsection{The Weyl tensor and the complexity factor}
 Some of the solutions exhibited in the next section are obtained from the condition of vanishing complexity factor.  This is  a scalar function intended to measure the degree of complexity of a given fluid distribution \cite{ps1, ps2}, and is related to the so called structure scalars \cite{sc}.
 
In  the spherically symmetric case the magnetic part of the Weyl tensor  ($C^{\rho}_{\alpha
\beta
\mu}$) vanishes, accordingly it is  defined by its ``electric'' part
 $E_{\gamma \nu }$, defined by 
   \begin{equation}
E_{\alpha \beta} = C_{\alpha \mu \beta \nu} V^\mu V^\nu,
\label{elec}
\end{equation}
whose non trivial components are
\begin{eqnarray}
E_{11}&=&\frac{2}{3}B^2 {\cal E},\nonumber \\
E_{22}&=&-\frac{1}{3} R^2 {\cal E}, \nonumber \\
E_{33}&=& E_{22} \sin^2{\theta},
\label{ecomp}
\end{eqnarray}
where
\begin{widetext}
\begin{eqnarray}
{\cal E}= \frac{1}{2 A^2}\left[\frac{\ddot R}{R} - \frac{\ddot B}{B} - \left(\frac{\dot R}{R} - \frac{\dot B}{B}\right)\left(\frac{\dot A}{A} + \frac{\dot R}{R}\right)\right]+ \frac{1}{2 B^2} \left[\frac{A^{\prime\prime}}{A} - \frac{R^{\prime\prime}}{R} + \left(\frac{B^{\prime}}{B} + \frac{R^{\prime}}{R}\right)\left(\frac{R^{\prime}}{R}-\frac{A^{\prime}}{A}\right)\right] - \frac{1}{2 R^2}.
\label{E}
\end{eqnarray}
\end{widetext}
 Observe that  the electric part of the
Weyl tensor, may be written as
\begin{equation}
E_{\alpha \beta}={\cal E} \left(K_\alpha K_\beta-\frac{1}{3}h_{\alpha \beta}\right).
\label{52}
\end{equation}
As shown in \cite{ps1, ps2} the complexity factor is identified with the scalar function $Y_{TF}$ which defines the trace--free part of the electric Riemann tensor (see \cite{sc} for details).

Thus,
let us define tensor $Y_{\alpha \beta}$ by
\begin{equation}
Y_{\alpha \beta}=R_{\alpha \gamma \beta \delta}V^\gamma V^\delta,
\label{electric}
\end{equation}

which may be expressed in terms of  two scalar functions $Y_T, Y_{TF}$, as
\begin{eqnarray}
Y_{\alpha\beta}=\frac{1}{3}Y_T h_{\alpha
\beta}+Y_{TF}\left(K_{\alpha} K_{\beta}-\frac{1}{3}h_{\alpha
\beta}\right).\label{electric'}
\end{eqnarray}

Then after lengthy but simple calculations, using field equations, we obtain (see \cite{1B} for details)
\begin{eqnarray}
Y_T=4\pi(\mu+3 P_r-2\Pi) , \qquad
Y_{TF}={\cal E}-4\pi \Pi. \label{EY}
\end{eqnarray}

Next, using  (\ref{12}), (\ref{14}), (\ref{15}) with (\ref{17masa}) and (\ref{E}) we obtain
\begin{equation}
\frac{3m}{R^3}=4\pi \left({\mu}-\Pi \right) - \cal{E},
\label{mE}
\end{equation}
which combined with (\ref{3m/R3})  and (\ref{EY}) produces

\begin{equation}
Y_{TF}= -8\pi\Pi +\frac{4\pi}{R^3}\int^r_0{R^3\left(D_R {\mu}-3{q}\frac{U}{RE}\right)R^\prime dr}.
\label{Y}
\end{equation}
It is worth noticing that due to a different signature, the sign of $Y_{TF}$ in the above equation differs from the sign of the $Y_{TF}$ used in \cite{ps1} for the static case.

Thus the scalar $Y_{TF}$ may be expressed through the Weyl tensor and the anisotropy of pressure  or in terms of the anisotropy of pressure, the density inhomogeneity and  the dissipative variables.

In terms of the metric functions the scalar $Y_{TF}$ reads

\begin{widetext}
\begin{eqnarray}
Y_{TF}= \frac{1}{A^2}\left[\frac{\ddot R}{R} - \frac{\ddot B}{B} + \frac{\dot A}{A}\left(\frac{\dot B}{B} - \frac{\dot R}{R}\right)\right]+ \frac{1}{ B^2} \left[\frac{A^{\prime\prime}}{A} -\frac{A^{\prime}}{A}\left(\frac{B^{\prime}}{B}+\frac{R^{\prime}}{R}\right)\right] .
\label{itfm}
\end{eqnarray}
\end{widetext}
\subsection{The exterior spacetime and junction conditions}
Since we are considering bounded fluid distributions then we still have to satisfy the junction (Darmois) conditions. Thus, outside $\Sigma$ we assume we have the Vaidya
spacetime (i.e.\ we assume all outgoing radiation is massless),
described by
\begin{equation}
ds^2=-\left[1-\frac{2M(v)}{\bf r}\right]dv^2-2d{\bf r}dv+{\bf r}^2(d\theta^2
+\sin^2\theta
d\phi^2) \label{1int},
\end{equation}
where $M(v)$  denotes the total mass,
and  $v$ is the retarded time.

The matching of the full nonadiabatic sphere   to
the Vaidya spacetime, on the surface $r=r_{\Sigma}=$ constant, requires the continuity of the first and second fundamental forms across $\Sigma$ (see \cite{chan} and references therein for details), which implies 
 \begin{equation}
m(t,r)\stackrel{\Sigma}{=}M(v), \label{junction1}
\end{equation}
and
\begin{widetext}
\begin{eqnarray}
2\left(\frac{{\dot R}^{\prime}}{R}-\frac{\dot B}{B}\frac{R^{\prime}}{R}-\frac{\dot R}{R}\frac{A^{\prime}}{A}\right)
\stackrel{\Sigma}{=}-\frac{B}{A}\left[2\frac{\ddot R}{R}
-\left(2\frac{\dot A}{A}
-\frac{\dot R}{R}\right)\frac{\dot R}{R}\right]+\frac{A}{B}\left[\left(2\frac{A^{\prime}}{A}
+\frac{R^{\prime}}{R}\right)\frac{R^{\prime}}{R}-\left(\frac{B}{R}\right)^2\right],
\label{j2}
\end{eqnarray}
\end{widetext}
where $\stackrel{\Sigma}{=}$ means that both sides of the equation
are evaluated on $\Sigma$.

Comparing (\ref{j2}) with  (\ref{13}) and (\ref{14}) one obtains
\begin{equation}
q\stackrel{\Sigma}{=}P_r.\label{j3}
\end{equation}
Thus   the smooth matching of
(\ref{1})  and (\ref{1int}) on $\Sigma$ implies (\ref{junction1}) and  (\ref{j3}).

Finally, the total luminosity ($L_\infty$) for an observer at rest at infinity is defined by 
\begin{equation}
L_\infty=4\pi P_{r\Sigma}\left[R^2\left(\frac{R^\prime}{B}+\frac{\dot R}{A}\right)^2 \right]_\Sigma.
\label{lum}
\end{equation}

\section{The transport equation}
In the dissipative case we shall need a transport equation in order to find  the temperature distribution and evolution. Assuming a causal dissipative theory (e.g.the Israel-- Stewart theory \cite{19nt, 20nt, 21nt} ) the transport equation for the heat flux reads
\begin{eqnarray}
\tau h^{\alpha \beta}V^\gamma q_{\beta;\gamma}&+&q^\alpha=-\kappa h^{\alpha \beta}\left(T_{,\beta}+Ta_\beta\right)\nonumber \\&-&\frac{1}{2}\kappa T^2 \left(\frac{\tau V^\beta}{\kappa T^2}\right)_{;\beta} q^\alpha,
\label{tre}
\end{eqnarray}
where $\kappa$ denotes the thermal conductivity, and $T$ and $\tau$ denote temperature and relaxation time respectively. 

In the spherically symmetric case under consideration, the transport equation has only one independent component which may be obtained from (\ref{tre}) by contracting with the unit spacelike vector $K^\alpha$, it reads
\begin{equation}
\tau V^\alpha  q_{,\alpha}+q=-\kappa \left(K^\alpha T_{,\alpha}+T a\right)-\frac{1}{2}\kappa T^2\left(\frac{\tau V^\alpha}{\kappa T^2}\right)_{;\alpha} q.
\label{5}
\end{equation}

Sometimes it is possible to simplify the equation above, in the so called truncated  transport equation, when the last term in (\ref{tre}) may be neglected \cite{t8}, producing 
\begin{equation}
\tau V^\alpha  q_{,\alpha}+q=-\kappa \left(K^\alpha T_{,\alpha}+T a\right).
\label{5trun}
\end{equation}
\section{The homologous and quasi--homologous conditions}

As mentioned before, in order to specify some of our models we shall impose the condition of vanishing complexity factor. However, for time dependent systems, it is not enough to define the complexity of the fluid distribution. We need also to elucidate  what is the simplest pattern of evolution of the system.

In \cite{ps2} the concept of  homologous evolution was introduced, in analogy with the same concept in classical astrophysics, as to represent the simplest mode of evolution of the fluid distribution. 

Thus,    the field equation (\ref{13}) written as
\begin{equation}
D_R\left(\frac{U}{R}\right)=\frac{4 \pi}{E} q+\frac{\sigma}{R},
\label{vel24}
\end{equation}
 can be easily integrated to obtain
\begin{equation}
U=\tilde a(t) R+R\int^r_0\left(\frac{4\pi}{E} q+\frac{\sigma}{R}\right)R^{\prime}dr,
\label{vel25}
\end{equation}
where $\tilde a$ is an integration function, or
\begin{equation}
U=\frac{U_\Sigma}{R_\Sigma}R-R\int^{r_\Sigma}_r\left(\frac{4\pi}{E} q+\frac{\sigma}{ R}\right)R^{\prime}dr.
\label{vel26}
\end{equation}
If the integral in the above equations vanishes  we have from (\ref{vel25}) or (\ref{vel26}) that
\begin{equation}
 U=\tilde a(t) R.
 \label{ven6}
 \end{equation}

 This relationship  is characteristic of the homologous evolution in Newtonian hydrodynamics \cite{20n, 21n, 22n}. In our case, this may occur if the fluid is shear--free and non dissipative, or if the two terms in the integral cancel each other.

In \cite{ps2}, the term  ``homologous evolution'' was used to characterize  relativistic systems satisfying, besides  (\ref{ven6}), the condition
\begin{equation}
\frac{R_1}{R_{2}}=\mbox{constant},
\label{vena}
\end{equation}
where $R_1$ and $R_{2}$ denote the areal radii of two concentric shells ($1,2$) described by $r=r_1={\rm constant}$, and $r=r_{2}={\rm constant}$, respectively.

The important point that we want to stress here is that (\ref{ven6}) does not imply (\ref{vena}).
Indeed, (\ref{ven6})  implies that for the two shells of fluids $1,2$ we have
\begin{equation}
\frac{U_1}{U_{2}}=\frac{A_{2} \dot R_1}{A_1 \dot R_{2}}=\frac{R_1}{R_{2}},
\label{ven3}
\end{equation}
that implies (\ref{vena}) only if  $A=A(t)$, which by a simple coordinate transformation becomes $A={\rm constant}$. Thus in the non--relativistic regime, (\ref{vena}) always follows from the condition that  the radial velocity is proportional to the radial distance, whereas in the relativistic regime the condition (\ref{ven6}) implies (\ref{vena}), only if the fluid is geodesic.

In \cite{epjc} the homologous condition was relaxed, leading to what was defined as  quasi--homologous evolution,  restricted only by condition  (\ref{ven6}), implying
\begin{equation}
\frac{4\pi}{R^\prime}B  q+\frac{\sigma}{ R}=0.
\label{ch1}
\end{equation}

\section{Conformal motions: exact solutions}
We shall consider spacetimes whose line element is defined by (\ref{1}), admitting a CKV, i.e. satisfying the equation
\begin{equation}
\mathcal{L}_\chi g_{\alpha \beta} =2\psi g_{\alpha \beta} \rightarrow \mathcal{L}_\chi g^{\alpha \beta} =-2\psi g^{\alpha \beta}, 
\label{1cmh}
\end{equation}
 where $\mathcal{L}_\chi$ denotes the Lie derivative with respect to the vector field ${\bold \chi}$, which unless specified otherwise, has the general form
\begin{equation}
\bold \chi=\xi(t, r, )\partial_t+\lambda(t, r, )\partial_r,
\label{2cmh}
\end{equation}
 and $\psi$ in principle is a function of $t, r$. The case $\psi=constant$ corresponds to a HKV.

Our goal consists in finding exact solutions  admitting a one parameter group of conformal motions, expressed in terms of elementary functions.

Two different families of solutions will be obtained depending on the choice of $ \chi^\alpha$. One of these families corresponds to the case   with $ \chi^\alpha$  orthogonal to $V^\alpha$, while the other corresponds to the case with $ \chi^\alpha$  parallel to $V^\alpha$. For both families  we shall consider separately the non--dissipative ($q=0$) and the dissipative ($q\neq$) case.

For the non--dissipative case of the family of solutions with $ \chi^\alpha$  orthogonal to $V^\alpha$, we shall obtain from the matching conditions and specific values of the relevant  parameters, solutions $I, II, III$, and for the particular case $M=0$ we shall obtain  solutions $IV, V$. For the dissipative case of this family, imposing the vanishing complexity factor condition and the shear--free condition we shall obtain  solution $VI$.

For the non--dissipative case of the family of solutions with $ \chi^\alpha$  parallel to $V^\alpha$, we shall obtain from the matching conditions and the vanishing complexity factor condition  solution $VII$, whereas from  specific values of relevant parameters we shall obtain solution $VIII$. Also imposing the condition $M=0$ we shall obtain in this case solutions $IX, X$.

Finally for the dissipative case of this family, imposing the complexity factor condition, we shall obtain solution $XI$.

Let us  start by considering the case $ \chi^\alpha$  orthogonal to $V^\alpha$ and $q=0$. 
\subsection{$\chi_\alpha V^\alpha=q=0$.}
Then from 
\begin{equation}
\mathcal{L}_\chi g_{\alpha \beta} =2\psi g_{\alpha \beta} =\chi^\rho \partial_\rho g_{\alpha \beta} +g_{\alpha \rho}\partial_\beta \chi^\rho+g_{\beta \rho}\partial_\alpha \chi^\rho,
\label{mod1}
\end{equation}
\\
we obtain
\begin{equation}
\psi=\chi^1\frac{A^\prime}{A},
\label{mod2}
\end{equation}
\begin{equation}
 \psi=\chi^1 \frac{B^\prime}{B}+(\chi^1)^\prime,
\label{mod3}
\end{equation}
\begin{equation}
\psi=\chi^1\frac{R^\prime}{R},
\label{mod4}
\end{equation}

and
\begin{equation}
\chi^1_{,t} =\chi^1_{,\theta}=\chi^1_{,\phi} =0.
\label{mod5}
\end{equation}

From  (\ref{mod2}) and (\ref{mod4}) it follows
\begin{equation}
A=h(t)R,
\label{mod6}
\end{equation}

where $h$ is an arbitrary function of $t$, which without loos of generality may be put equal to $1$ by reparametrizing $t$.

Thus we may write
\begin{equation}
A=\alpha R,
\label{mod7a}
\end{equation}
where $\alpha$ is a unit constant with dimensions of $\frac{1}{length}$.

Next, taking the time derivative of (\ref{mod3}) and (\ref{mod4}) and using (\ref{mod5}) we obtain
\begin{equation}
\frac{B}{G(r)}=\alpha F_1(t) R,
\label{mod8a}
\end{equation}
where $G(r)$ is an arbitrary function of $r$  which may be put equal to $1$ by a a reparametrization  of $r$, and $F_1$ is an arbitrary dimensionless function of $t$.

Thus we have 
\begin{equation}
B=\alpha F_1(t) R,
\label{mod9a}
\end{equation}
and 
\begin{equation}
(\chi^1)^\prime=0\Rightarrow \chi^1=constant.
\label{mod10}
\end{equation}

Then, feeding back (\ref{mod7a}) and (\ref{mod9a}) into (\ref{13}) with $q=0$, one obtains
\begin{equation}
A=\alpha R=\frac{F(t)}{f(t)+g(r)},\quad B=\frac{1}{f(t)+g(r)},
\label{mod11}
\end{equation}
where $f$ and $g$ are two arbitrary functions of their arguments and $F(t)\equiv \frac{1}{F_1(t)}$.

So far we see that any model is determined up to three arbitrary functions $F(t), f(t), g(r)$.

Then the field equations read 
\begin{widetext}
\begin{equation}
8\pi \mu=\frac{(f+g)^2}{F^2}\left[\frac{\dot F^2}{F^2}-\frac{4\dot F \dot f}{F(f+g)}+\frac{3\dot f^2}{(f+g)^2}\right]+2g^{\prime \prime}(f+g)-3g^{\prime 2}+\frac{\alpha^2 (f+g)^2}{F^2},
\label{mod12}
\end{equation}
\end{widetext}

\begin{widetext}
\begin{equation}
8\pi P_r=\frac{(f+g)^2}{F^2}\left[\frac{\dot F^2}{F^2}+\frac{2\dot F \dot f}{F(f+g)}-\frac{3\dot f^2}{(f+g)^2}+\frac{2\ddot f}{f+g}-\frac{2\ddot F}{F}\right]+3g^{\prime 2}-\frac{\alpha^2 (f+g)^2}{F^2},
\label{mod13}
\end{equation}
\end{widetext}

\begin{widetext}
\begin{equation}
8\pi P_\bot=\frac{(f+g)^2}{F^2}\left[\frac{\dot F^2}{F^2}-\frac{3\dot f^2}{(f+g)^2}+\frac{2\ddot f}{f+g}-\frac{\ddot F}{F}\right]+3g^{\prime 2}-2g^{\prime \prime} (f+g).
\label{mod14}
\end{equation} 
\end{widetext}

Using  the results above the matching conditions (\ref{junction1}) and (\ref{j3}) on the surface $r=r_\Sigma=constant$  read

\begin{equation}
\dot R_\Sigma^2+\alpha^2(R_\Sigma^2-2MR_\Sigma-\omega R_\Sigma ^4)=0,
\label{mod15}
\end{equation}

and

\begin{equation}
2\ddot R_\Sigma R_\Sigma  -\dot R^2_\Sigma-\alpha^2(3\omega R_\Sigma ^4-R_\Sigma ^2)=0,
\label{mod16}
\end{equation}
with $\omega\equiv g^\prime(r_\Sigma)^2$.

\begin{figure}[h]
\includegraphics[width=3.in,height=4.in,angle=0]{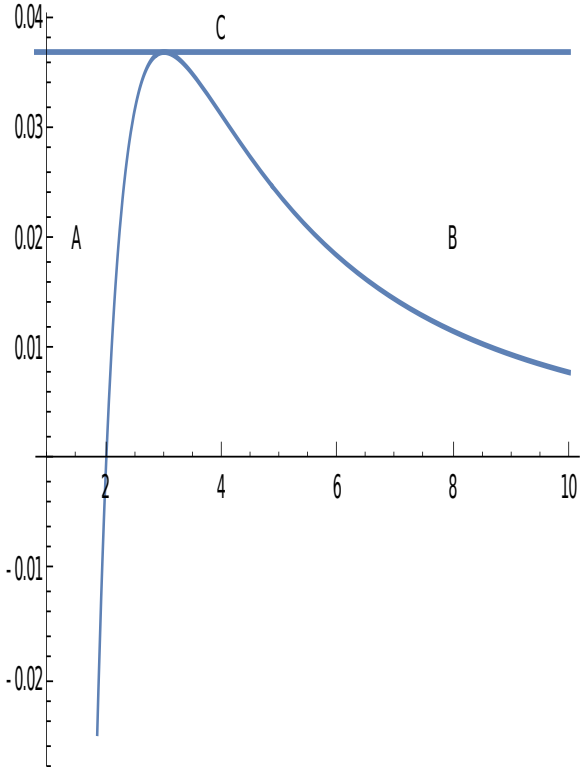}
\caption {\it  $\tilde V$ as function of $x$.  The horizontal line corresponds to the value of $\tilde V=1/27$}
\end{figure}

It is a simple matter to check that (\ref{mod15}) is just the first integral of (\ref{mod16}), therefore we only need to consider the former equation.

It would be useful to write (\ref{mod15}) in the form

\begin{equation}
\dot R^2_\Sigma=\alpha^2R^4_\Sigma \left[\omega-V(R_\Sigma)\right],
\label{mod17}
\end{equation}
with
\begin{equation}
V(R_\Sigma)=\frac{1}{R_\Sigma^2}-\frac{2M}{R_\Sigma^3},
\label{mod18}
\end{equation}
or 

\begin{equation}
\tilde V(R_\Sigma)\equiv M^2 V(R_\Sigma)=\frac{1}{x^2}-\frac{2}{x^3},
\label{mod19}
\end{equation}
with $x\equiv \frac{R_\Sigma}{M}$.

The maximum of $\tilde V$  ($\tilde V_{max.}=\frac{1}{27}$) occurs at $x=3$ ($R_\Sigma=3M)$, whereas  $\tilde V$ vanishes at at $x=2$ ($R_\Sigma=2M)$.

Obviously  all solutions have to satisfy the conditions $\omega>V(R_\Sigma)$. Among them we have:
\begin{itemize}
\item Solutions with $\omega<\frac{1}{27M^2}$. In this case we may have solutions evolving   between the singularity and some value of $R_\Sigma$ in the interval $[2M,3M]$ (region A in figure 1), and solutions with $R_\Sigma$ in the interval $[3M, \infty]$ (region B in figure 1).
\item Solutions with $\omega>\frac{1}{27M^2}$ in which case $R_\Sigma$ is in the interval $[0, \infty]$ (region C in figure 1).
\item Solutions with $\omega=\frac{1}{27M^2}$ in which case $R_\Sigma$ may be  in the interval $[0, 3M]$ or in the interval $[3M, \infty]$

\item Solutions with $\omega=0$. In which case $R_\Sigma$  oscillates in the interval $[0, 2M]$.
\end{itemize}


In general, we may write from (\ref{mod17}) 
\begin{equation}
 \alpha(t-t_0)=\pm\int{\frac{dR_\Sigma}{R_\Sigma^2 \sqrt{\omega-V(R_\Sigma)}}},
\label{mod20}
\end{equation}
from which we may obtain $R_\Sigma$ expressed in terms of elliptic  functions. However, in some cases analytical solutions may be found in terms of elementary functions. For doing that we shall proceed as follows.

\noindent Let us introduce the variable  $r=x+b$  in the polynomial

\begin{equation}
P(r)=a_0 r^4+a_1 r^3+a_2r^2+a_3r+a_4,
\end{equation}

\noindent which allows us to write
\begin{widetext}
\begin{eqnarray}
  P(x) &=& a_0 x^4+(a_1+4a_0b)x^3+(a_2+3a_1b+6a_0b^2)x^2
   + (a_3+2a_2b+3a_1b^2+4a_0b^3)x \nonumber \\&+&a_4+a_3b+a_2b^2+a_1b^3+a_0b^4,
\end{eqnarray}
\end{widetext}

\noindent or
\begin{equation}\label{polf}
  P(x)=(\sqrt{a_0}x+b_1)^2(x+b_2)(x+b_3),
\end{equation}

\noindent where $b_{1,2,3}$  and  $b$ are solutions of the following equations
\begin{equation}
  a_4+a_3b+a_2b^2+a_1b^3+a_0b^4= b_1^2 b_2 b_3,\label{ec1} 
  \end{equation}
\begin{equation}
a_3+2a_2b+3a_1b^2+4a_0b^3=b_1^2(b_2+b_3)+2\sqrt{a_0}b_1 b_2 b_3,
  \end{equation}
\begin{equation}
  a_2+3a_1b+6a_0b^2= b_1^2+2\sqrt{a_0}b_1(b_2+b_3)+a_0 b_2 b_3,
  \end{equation}
\begin{equation}
  a_1+4a_0b =2\sqrt{a_0}b_1+a_0(b_2+b_3).\label{ec4}
  \end{equation}

\noindent  Then the integration of (\ref{mod20}) produces
\begin{widetext}
\begin{equation}\label{int}
  \pm \int \frac{dx}{(\sqrt{a_0 }x+b_1)\sqrt{(x+b_2)(x+b_3)}}=\left\{\beta_1 \arctan \left(\beta_2\sqrt{\frac{b_2+x}{b_3+x}}\right) \atop \beta_1 \arctan^{-1}\left(\beta_2\sqrt{\frac{b_2+x}{b_3+x}}\right), \right.
\end{equation}
\end{widetext}
\noindent where
\begin{eqnarray}
\beta_1 &=& \frac{2}{\sqrt{(b_1-\sqrt{a_0}b_2)(\sqrt{a_0}b_3-b_1)}}, \\
  \beta_2 &=& \sqrt{\frac{\sqrt{a_0}b_3-b_1}{b_1-\sqrt{a_0}b_2}}.
\end{eqnarray}
To obtain  explicit solutions expressed through elementary functions, we shall assume $b=0$, thus in our notation 
 we have

\begin{eqnarray} \label{vc}
  a_0 &=& \omega, \\
  a_1 &=& a_4=0, \\
  a_2 &=& -1\qquad a_3=2M. \label{vc1}
\end{eqnarray}

\noindent Imposing  $b=0$, we are led  to two sub-cases, $b_2=0$, or  $b_3=0$, in both sub--cases $M=\frac{1}{3\sqrt{3\omega}}$. 

Using (\ref{vc})--(\ref{vc1}) in (\ref{int}), we obtain for both sub--cases the same solutions, namely

  \begin{equation}\label{caso1}
    R^{(I)}_\Sigma=\frac{6 M\tanh ^2\left[\frac{\alpha}{2}(t-t_0)\right]}{3-\tanh ^2\left[\frac{\alpha}{2}(t-t_0)\right]},
  \end{equation}

and 

\begin{equation}\label{caso1ba}
    R^{(II)}_\Sigma=\frac{6 M\coth^2\left[\frac{\alpha}{2}(t-t_0)\right]}{3-\coth^2\left[\frac{\alpha}{2}(t-t_0)\right]}.
  \end{equation}


In the first case, the areal radius of the boundary ($R^{(I)}_\Sigma$) expands from $0$ (the singularity) approaching $3M$ asymptotically as $t \rightarrow\infty$, thereby representing a white hole scenario.

In the second case, the areal radius of the boundary ($R^{(II)}_\Sigma$) contracts from $\infty$ (for $\coth^2\left[\frac{\alpha}{2}(t-t_0)\right]=3)$ approaching $3M$ asymptotically as $t\rightarrow\infty$.

Thus the fulfillment of the matching conditions provides  one of the arbitrary functions of time describing our metric. In order to specify further our model we shall impose the quasi--homologous evolution and the vanishing complexity factor condition.


\noindent As we can see from (\ref{vel26}), in the non--dissipative case the quasi--homologous condition implies that the fluid is shear--free ($\sigma=0$), implying in its turn

\begin{equation}\label{csigma}
  \frac{\dot{B}}{B}=\frac{\dot{R}}{R}\,\,\Rightarrow\,\, F(t)=Const.\equiv F_0.
\end{equation}

\noindent Thus the metric functions become
\begin{equation}\label{fmetricas}
  A=\frac{F_0}{f(t)+g(r)},\quad B=\frac{1}{f(t)+g(r)},\quad R=\frac{F_0}{\alpha\left[f(t)+g(r)\right]}.
\end{equation}

Therefore our models are now specified up to an arbitrary function of $r$ ($g(r)$). In order to fix this function we shall further impose the vanishing complexity factor condition.

\noindent Then feeding back (\ref{fmetricas}) into (\ref{itfm}) we obtain 
\begin{equation}\label{YTF}
  \frac{A^{\prime\prime}}{A}-\frac{A^{\prime}}{A}(\frac{B^{\prime}}{B}+\frac{R^{\prime}}{R})=0.
\end{equation}

\noindent Using  (\ref{fmetricas}) in  (\ref{YTF}), it follows at once

\begin{equation}\label{ecg}
  g(r)=c_1 r+c_2,
\end{equation}

\noindent  with  $c_1\equiv \pm\sqrt{\omega}$, and   $c_2$ is another integration constant, we shall choose the negative sign in $c_1$  in order to ensure that $R^\prime>0$. However it should be noticed that  the regularity conditions, necessary to ensure elementary flatness in the vicinity of  the axis of symmetry, and in particular at the center (see \cite{1n}, \cite{2n}, \cite{3n}), are not satisfied.

Therefore after the imposition of the two conditions above (quasi--homologous evolution and vanishing complexity factor) we have all the metric  functions completely specified for any of the above solutions to (\ref{mod17}). 

Thus in the case $b=0$ we obtain from (\ref{caso1}) 

\begin{equation}\label{fdt}
  f^{(I)}(t)=\frac{F_0\left\{3-\tanh^2\left[\frac{\alpha}{2}(t-t_0)\right]\right\}}{6\alpha M \tanh^2\left[\frac{\alpha}{2}(t-t_0)\right]}-c_1r_\Sigma-c_2,
\end{equation}

\noindent from which the physical variables are easily found to  be 
\begin{widetext}
\begin{eqnarray}
 8\pi\mu &=& -3c_1^2+\frac{\left\{-F_0-6\alpha c_1M(r_\Sigma-r)+3F_0\coth^2[\frac{\alpha}{2}(t-t_0)]\right\}^2}{36F_0^2M^2}
          +\frac{3\cosh^2[\frac{\alpha}{2}(t-t_0)]}{4M^2 \sinh^6[\frac{\alpha}{2}(t-t_0)]} ,\\
8\pi P_r &=& \frac{2\alpha c_1(r_\Sigma-r)}{3F_0M}
            -\frac{\alpha^2c_1^2 (r_\Sigma-r)^2}{F_0^2}-\frac{3c_1\alpha (r_\Sigma-r)}{2F_0M\sinh^4[\frac{\alpha}{2}(t-t_0)]},\label{pr0a} \\
  8\pi P_\bot&=& 3c_1^2+\frac{\left[F_0-3\alpha c_1M(r_\Sigma-r)\right]}{3F_0M^2\sinh^2[\frac{\alpha}{2}(t-t_0)]}
  +\frac{\left[F_0-6\alpha c_1M(r_\Sigma-r)\right]}{4F_0M^2\sinh^4[\frac{\alpha}{2}(t-t_0)]}.
\end{eqnarray}
\end{widetext}
From (\ref{pr0a})  it follows at once that $(P_r)_\Sigma=0$.

It is worth noticing that  the expansion scalar for this model reads
\begin{equation}
\Theta=\frac{3 \cosh[\frac{\alpha}{2}(t-t_0)]}{2M \sinh^3[\frac{\alpha}{2}(t-t_0)]}.
\label{theta1}
\end{equation}

Thus the expansion is homogeneous and positive, diverging at $t=t_0$, and tending to zero as $t\rightarrow\infty$. The fast braking of the expansion for $t>t_0$ is produced by the negative  initially large (diverging at $t=t_0$)  value of $D_TU$. This can be checked from (\ref{3m}), where the negative gravitational term proportional to $m/R^2$  provides the leading term in the equation ($\frac{m}{R^2}\sim \frac{1}{\sinh^4[\frac{\alpha}{2}(t-t_0)]})$. As time goes on  there is a sharp decreasing in the inertial mass density ($\mu+P_r$) which as  $t\rightarrow\infty$  becomes arbitrarily small (see (\ref{sl}) below).   Now, the striking fact is that the equilibrium is reached asymptotically, but not, as usual,  by the balance between   the gravitational  term (the first term on the right of (\ref{3m})) and the hydrodynamic terms (the second term on the right of (\ref{3m})). Instead, both terms cancel independently. Indeed, as $t\rightarrow\infty$, the gravitational term vanishes due to the fact that the inertial mass density (the ``passive gravitational mass density'')  $\mu+P_r\rightarrow 0$, and the hydrodynamic term vanishes because, as it can be easily checked, the radial pressure gradient cancels the anisotropic factor, as $t\rightarrow\infty$.

Next, if we take (\ref{caso1ba}) we obtain for $f(t)$
\begin{equation}\label{fdt}
  f^{(II)}(t)=\frac{F_0\left\{3-\coth^2\left[\frac{\alpha}{2}(t-t_0)\right]\right\}}{6\alpha M \coth^2\left[\frac{\alpha}{2}(t-t_0)\right]}-c_1r_\Sigma-c_2,
\end{equation}

\noindent whereas the expressions for the physical variables read
\begin{widetext}
\begin{eqnarray}
8\pi\mu &= &-3c_1^2 +\frac{3\sinh^2[\frac{\alpha}{2}(t-t_0)]}{4M^2 \cosh^6[(\frac{\alpha}{2}(t-t_0)]} +\frac{\left\{F_0+6\alpha c_1M(r_\Sigma-r)-3F_0\tanh^2[\frac{\alpha}{2}(t-t_0)]\right\}^2}{36F_0^2M^2},
 \\
  8\pi P_r &=& \frac{2\alpha c_1(r_\Sigma-r)}{3F_0M} -\frac{\alpha^2 c_1^2(r_\Sigma-r)^2}{F_0^2}
  -\frac{3c_1\alpha (r_\Sigma-r)}{2F_0M\cosh^4[\frac{\alpha}{2}(t-t_0)]}, \\
  8\pi P_\bot&=& 3c_1^2-\frac{F_0-3\alpha c_1 M(r_\Sigma-r)}{3F_0M^2 \cosh^2[\frac{\alpha}{2}(t-t_0)]}
  +\frac{F_0-6\alpha c_1 M(r_\Sigma-r)}{4F_0M^2 \cosh^4[\frac{\alpha}{2}(t-t_0)]}.
\end{eqnarray}
\end{widetext}

In the limit $t\rightarrow\infty$ the two above solutions converge to the same  static distribution  whose physical variables are
\begin{eqnarray}\label{sl}
 8\pi\mu = \frac{1}{9M^2}\left[3+\left(\frac{r}{r_\Sigma}\right)^2-4\left(\frac{r}{r_\Sigma}\right)\right] ,\\\nonumber
8\pi P_r = \frac{1}{9M^2}\left[-3-\left(\frac{r}{r_\Sigma}\right)^2+4\left(\frac{r}{r_\Sigma}\right)\right],\\ \nonumber
  8\pi P_\bot= \frac{1}{9M^2},
\end{eqnarray}
where  the constant $F_0$ has been chosen $F_0=\frac{r_\Sigma \alpha}{\sqrt{3}}$. It is worth noticing that the ensuing equation of state for the static limit is the Chaplygin--type equation $\mu=-P_r$.

In the case  $\omega=0$ the expression for $R_\Sigma$ is given by  
\begin{equation}
R^{(III)}_\Sigma=2M\cos^2\frac{\alpha}{2}(t-t_0),
\label{mod23}
\end{equation}

and from(\ref{ecg}) 
\begin{equation}
\omega=g^\prime (r_\Sigma)^2=c_1^2=0.
\end{equation}

\noindent    Then the following expressions may be obtained for the metric functions

\begin{equation}\label{fmetricasI}
  A=\frac{F_0}{f(t)+c_2},\qquad B=\frac{1}{f(t)+c_2}\qquad R=\frac{F_0}{\alpha\left[f(t)+c_2\right]},
\end{equation}

\noindent with   $f(t)\equiv f^{III}(t)$, given by 

\begin{equation}\label{fdt2}
  f^{(III)}(t)=\frac{F_0}{2\alpha M \cos^2 \left[\frac{\alpha}{2}(t-t_0)\right]}-c_2,
\end{equation}

\noindent and the physical variables read

\begin{eqnarray}
  8\pi\mu &=& \frac{\left\{3-2\cos^2[\frac{\alpha}{2}(t-t_0)]\right\}}{4 M^2 \cos^6[\frac{\alpha}{2}(t-t_0)]},\\
  8\pi P_r &=& 0, \\
  8\pi P_\bot&=& \frac{1}{4M^2 \cos^4[\frac{\alpha}{2}(t-t_0)]}.
\end{eqnarray}

It is worth stressing the presence of important  topological pathologies in this solution (e.g. $R^\prime=0$), implying the appearance of shell crossing singularities.

Before closing  this subsection we would like to call the attention to a very peculiar solution that may be obtained by assuming that the space--time outside the boundary surface delimiting the fluid is Minkowski. This implies $M=0$, and then the solutions to (\ref{mod15}) read
\begin{equation}
R^{(IV)}_\Sigma=\frac{1}{\sqrt{\omega} \cos[\alpha(t-t_0)]},
\label{gos1}
\end{equation}

and
\begin{equation}
R^{(V)}_\Sigma=\frac{1}{\sqrt{\omega} \sin[\alpha(t-t_0)]}.
\label{gos1b}
\end{equation}

Assuming further that the evolution is quasi--homologous and the complexity factor vanishes, we obtain  for the functions $f(t)$
\begin{equation}
f^{(IV)}(t)=\frac{1}{\alpha}F_0\sqrt{\omega} \cos[\alpha(t-t_0)]+\sqrt{\omega}r_\Sigma-c_2,
\label{gos2}
\end{equation}
and 
\begin{equation}
f^{(V)}(t)=\frac{1}{\alpha}F_0\sqrt{\omega} \sin[\alpha(t-t_0)]+\sqrt{\omega}r_\Sigma-c_2.
\label{gos2b}
\end{equation}

The corresponding physical variables for $f^{(IV)}$ read
\begin{eqnarray}
8\pi \mu&=&-2\omega \cos^2[\alpha(t-t_0)] \nonumber \\&+&\omega(1-\frac{r}{r_\Sigma})\left[1-\frac{r}{r_\Sigma}+2\cos[\alpha(t-t_0)]\right],
\label{mu03}
\end{eqnarray}
\begin{equation}
8\pi P_r=-\omega \left(1-\frac{r}{r_\Sigma}\right)\left[ 4 \cos[\alpha(t-t_0)]+1-\frac{r}{r_\Sigma}\right],
\label{mu03b}
\end{equation}

\begin{equation}
8\pi P_\bot=\omega \cos^2[\alpha(t-t_0)]-2\omega\left(1-\frac{r}{r_\Sigma}\right)\cos[\alpha(t-t_0)],
\label{mu03c}
\end{equation}

whereas for $f^{(V)}$ they are

\begin{eqnarray}
8\pi \mu&=&-2\omega \sin^2[\alpha(t-t_0)]\nonumber \\&+&\omega(1-\frac{r}{r_\Sigma})\left[1-\frac{r}{r_\Sigma}+2\sin[\alpha(t-t_0)]\right],
\label{mu03d}
\end{eqnarray}
\begin{equation}
8\pi P_r=-\omega \left(1-\frac{r}{r_\Sigma}\right) \left [4\sin[\alpha(t-t_0)]+1-\frac{r}{r_\Sigma}\right],
\label{mu03e}
\end{equation}

\begin{equation}
8\pi P_\bot=\omega \sin^2[\alpha(t-t_0)]-2\omega\left(1-\frac{r}{r_\Sigma}\right)\sin[\alpha(t-t_0)].
\label{mu03f}
\end{equation}

In the above the constants $F_0, \alpha, r_\Sigma$ have been chosen such that  $\frac{F_0}{\alpha r_\Sigma}=1$.

This kind of configurations have been considered in \cite{11co, zel}.
\subsection{$\chi_\alpha V^\alpha=0; q\neq0$}
Let us now consider the general dissipative case when the vector $\chi^\alpha$ is orthogonal to the four--velocity.  

Then from (\ref{mod1})  we obtain, following the same procedure as in the non--dissipative case

\begin{equation}
A=\alpha R,
\label{mod7}
\end{equation}
where $\alpha$ is a unit constant with dimensions of $\frac{1}{length}$,

\begin{equation}
F(t) B=\alpha  R,
\label{mod9}
\end{equation}
where $F(t)$ is and arbitrary function of $t$,
and 
\begin{equation}
\chi^1_{,1}=0\Rightarrow \chi^1=constant.
\label{mod10}
\end{equation}

Then, feeding back (\ref{mod7}) and (\ref{mod9}) into (\ref{13}) with $q\neq0$, one obtains
\begin{equation}
\frac{\dot B^\prime}{B}-\frac{2\dot B B^\prime}{B^2}=4\pi qAB.
\label{dis1}
\end{equation}

The equation above may be formally integrated, to obtain
\begin{equation}
B=\frac{1}{f(t)+g(r)-4\pi \int \int q Adt dr},
\label{dis2}
\end{equation}

\begin{equation}
A=\alpha R=\frac{F(t)}{f(t)+g(r)-4 \pi \int \int q Adt dr},
\label{mod11}
\end{equation}
where $f$ and $g$ are two arbitrary functions of their arguments.

In order to find a specific solutions we shall impose next the  vanishing complexity factor condition ($Y_{TF}=0$).

Then  from the above expressions and (\ref{itfm}), the condition $Y_{TF}=0$ reads
\begin{equation}
\frac{1}{F^2}\left(\frac{\ddot F}{F}+\frac{\dot F \dot B}{FB}-\frac{\dot F^2}{F^2}\right)+\frac{B^{\prime \prime}}{B}-2\left(\frac{B^\prime}{B}\right)^2=0.
\label{dis4}
\end{equation}

In order to find a solution to the above equation we shall assume that 
\begin{equation}
\left(\frac{\ddot F}{F}+\frac{\dot F \dot B}{FB}-\frac{\dot F^2}{F^2}\right)=0,
\label{dis5}
\end{equation}
and
\begin{equation}
\frac{B^{\prime \prime}}{B}-2\left(\frac{B^\prime}{B}\right)^2=0.
\label{dis5b}
\end{equation}

The integration of (\ref{dis5b})  produces
\begin{equation}
B=-\frac{1}{\beta(t) r+\gamma(t)},
\label{dis6}
\end{equation}
where $\beta, \gamma$ are arbitrary functions of $t$. It is worth noticing that $\beta$ has dimensions of $\frac{1}{length}$, and $\gamma$ is dimensionless.

Next, taking the $r$ derivative of (\ref{dis5}) we obtain $\gamma=\frac{\beta}{\alpha}$.

Then  we may write
\begin{equation}
B=-\frac{\alpha}{\beta(t) (\alpha r+1)}.
\label{dis7}
\end{equation}

Next, combining (\ref{dis5}) with (\ref{dis7}) we obtain 
\begin{equation}
F(t)=c_3 e^{c_4\int \beta dt},
\label{dis7b}
\end{equation}
where $c_3, c_4$ are arbitrary constants.

From the above expression it follows at once
\begin{equation}
\frac{\dot F}{F}=c_4\beta.
\label{dis7c}
\end{equation}

On the other hand, (\ref{dis2}) with (\ref{dis7}), imply

\begin{equation}
4\pi \int \int q Adt dr=\beta(t) r, \quad g(r)=g_1 r+g_0,
\label{dis8}
\end{equation}

where $g_1, g_0$ are constant.

We may now write the physical variables in terms of the function $\beta(t)$, they read

\begin{equation}
8\pi \mu=\frac{\beta^2(\alpha r +1)^2}{\alpha^2 c_3^2e^{2c_4\int{\beta dt}}}\left(\alpha^2-4c_4\dot \beta+ \frac{3 \dot \beta^2}{\beta^2}+c_4^2\beta^2 \right)-3\beta^2,
\label{muidsco}
\end{equation}

\begin{equation}
4\pi q=-\frac{\beta \dot \beta (\alpha r+1)}{\alpha c_3 e^{c_4\int{\beta dt}}},
\label{qdiscom}
\end{equation}

\begin{equation}
8\pi P_r=-\frac{\beta^2(\alpha r +1)^2}{\alpha^2 c_3^2e^{2c_4\int{\beta dt}}}\left(\alpha^2-\frac{2\ddot \beta}{\beta}+ \frac{3 \dot \beta^2}{\beta^2}+c_4^2\beta^2 \right)+3\beta^2,
\label{pridsco}
\end{equation}

\begin{equation}
8\pi P_\bot=-\frac{\beta^2(\alpha r +1)^2}{\alpha^2 c_3^2e^{2c_4\int{\beta dt}}}\left(c_4\dot \beta-\frac{2\ddot \beta}{\beta}+ \frac{3 \dot \beta^2}{\beta^2}\right)+3\beta^2.
\label{ptidsco}
\end{equation}

The function $\beta$  may be found, in principle, from the junction condition (\ref{j3}), however since this is  in practice quite difficult at this level of generality, we shall first  impose further  constraints on our fluid distribution in order to obtain  a simpler model, and afterwards we shall use the junction conditions.

We shall  start  by imposing the quasi--homologous condition (\ref{ch1}). Then using (\ref{dis2}) and (\ref{mod11}) in (\ref{ch1}) we get
\begin{equation}
4\pi q A\frac{B^2}{B^\prime}=\frac{\dot F}{F}.
\label{dis3}
\end{equation} 

Using (\ref{dis3}) with (\ref{dis7})--(\ref{dis8}) one obtains
\begin{equation}
\beta=-\frac{1}{c_4 t+c_5},
\label{dis8c}
\end{equation}
where $c_5$ is a constant with dimensions of $length$.

So the metric functions may be written as

\begin{equation}
B=\frac{\alpha (c_4 t+c_5)}{( \alpha r+1)},
\label{dis9}
\end{equation}
\begin{equation}
A=\alpha R=\frac{c_3}{\alpha r+1}.
\label{mod11b}
\end{equation}
It is worth noticing that  the areal radius is independent on time ($U=0$), solutions of this kind have been found in \cite{epjc}

Next, instead of quasi--homologous condition we shall impose  the shear--free condition.  Then assuming $\sigma =0$ it follows at once that  $\dot F=0$ implying $c_4=0$. Then the metric functions become
\begin{equation}
B=-\frac{\alpha}{\beta(\alpha r+1)},\quad A=-\frac{c_3\alpha}{\beta(\alpha r+1)},\quad R=-\frac{c_3}{\beta(\alpha r+1)},
\label{metricsf1}
\end{equation}
from which we can write  the physical variables as
\begin{equation}
8\pi \mu=\frac{\beta^2(\alpha r +1)^2}{\alpha^2 c_3^2}\left(\alpha^2+ \frac{3 \dot \beta^2}{\beta^2}\right)-3\beta^2,
\label{muidscosf}
\end{equation}

\begin{equation}
4\pi q=-\frac{\beta \dot \beta (\alpha r+1)}{\alpha c_3},
\label{qdiscomsf}
\end{equation}

\begin{equation}
8\pi P_r=-\frac{\beta^2(\alpha r +1)^2}{\alpha^2 c_3^2}\left(\alpha^2-\frac{2\ddot \beta}{\beta}+ \frac{3 \dot \beta^2}{\beta^2} \right)+3\beta^2,
\label{pridscosf}
\end{equation}

\begin{equation}
8\pi P_\bot=-\frac{\beta^2(\alpha r +1)^2}{\alpha^2 c_3^2}\left(-\frac{2\ddot \beta}{\beta}+ \frac{3 \dot \beta^2}{\beta^2}\right)+3\beta^2.
\label{ptidscosf}
\end{equation}

Now we can find $\beta$ from the junction condition (\ref{j3}), which using (\ref{qdiscomsf}) and (\ref{pridscosf}) reads
\begin{equation}
\frac{2\ddot \beta}{\beta}-3\left(\frac{\dot \beta}{\beta}\right)^2+\frac{2\alpha_1 \dot \beta}{\beta}=\alpha^2-3\alpha^2_1,
\label{sol1}
\end{equation}
with
\begin{equation}
\alpha_1\equiv\frac{\alpha c_3}{\alpha r_\Sigma+1}.
\label{sol2}
\end{equation}

In order to integrate the above equation, let us introduce the variable $s=\frac{\dot \beta}{\beta}$, which casts (\ref{sol1}) into the Ricatti equation
\begin{equation}
2\dot s-s^2+2\alpha_1s=\alpha^2-3\alpha^2_1,
\label{sol3}
\end{equation}
whose solution is
\begin{equation}
s=\alpha_1+\sqrt{\alpha^2-4\alpha^2_1}\tan\left[\frac{\sqrt{\alpha^2-4\alpha^2_1}}{2}(t-t_0)\right],
\label{sol4}
\end{equation}
producing for $\beta$
\begin{equation}
\beta(t)=\alpha_2 e^{\alpha_1 t}\sec^2\left[\frac{\sqrt{\alpha^2-4\alpha^2_1}}{2}(t-t_0)\right],
\label{sol5}
\end{equation}
where $\alpha_2$ is a negative constant of integration with the same dimensions as $\beta$.

Using the truncated version of the transport equation (\ref{5trun}), we obtain for the temperature
\begin{equation}
T(t,r)= \frac{\alpha r+1}{4\pi c_3 \kappa \alpha}\left[\frac{\tau r (\dot \beta^2+\beta \ddot \beta)}{c_3}-\dot\beta \ln(\alpha r+1) \right]+T_0(t),
\label{tem}
\end{equation}
where $c_3$ and $T_0(t)$ are arbitrary constant  and function of integration, respectively. The model described by  equations (\ref{metricsf1})--(\ref{ptidscosf}) and  (\ref{sol5}), (\ref{tem}) will be named as model $VI$.

\subsection{$q=0; \chi^\alpha \vert\vert V^\alpha$}
We shall next analyze the case when the vector ${\bold \chi}$ is parallel to the four--velocity vector. We start by considering the non--dissipative case.
In this case the equation (\ref{mod1}) produces 
\begin{equation}
A=Bh(r), \qquad R=Br, \qquad \chi^0=1, \qquad \psi=\frac{\dot B}{B},
\label{pa1}
\end{equation}
where $h(r)$ is an arbitrary function of its argument. It is worth noticing that in this case the fluid is necessarily shear--free.

Thus the line element may be written as
\begin{equation}
ds^2=B^2\left[-h^2(r) dt^2+dr^2+r^2(d\theta^2+\sin^2\theta d\phi^2)\right].
\label{pa2}
\end{equation}

Next, using (\ref{pa1}) in (\ref{13}), the condition $q=0$ reads
\begin{equation}
\frac{\dot R^\prime}{R}-2\frac{\dot R}{R}\frac{R^\prime}{R}-\frac{\dot R}{R}\left(\frac{h^\prime}{h}-\frac{1}{r}\right)=0,
\label{pa3}
\end{equation}
whose solution is
\begin{eqnarray}
R=\frac{r}{h(r)\left[f(t)+g(r)\right]},
\label{pa3bt}
\end{eqnarray}
implying
\begin{eqnarray}
B=\frac{1}{h(r)\left[f(t)+g(r)\right]},\quad A=\frac{1}{f(t)+g(r)}, 
\label{pa3btb}
\end{eqnarray}

where $g, f$ are two arbitrary functions of their argument.

Thus the metric is defined up to three arbitrary functions ($g(r), f(t), h(r)$).

The function $f(t)$ will be obtained from the junction conditions (\ref{junction1}), (\ref{j3}).

Indeed, evaluating the mass function at the boundary surface $\Sigma$ we obtain from (\ref{junction1}) and (\ref{pa3bt})
\begin{equation}
\dot R^2_\Sigma= \alpha^2 R^4_\Sigma \left[\epsilon-V(R_\Sigma)\right],
\label{ppar1}
\end{equation}
where $\alpha^2\equiv \frac{h^2_\Sigma}{r^2_\Sigma}$, $\epsilon\equiv (g^\prime)^2_\Sigma h^2_\Sigma$, and 
\begin{equation}
V(R_\Sigma)=\frac{2 \sqrt{\epsilon}}{R_\Sigma}(1-a_1)+\frac{a_1}{R^2_\Sigma} (2-a_1)-\frac{2M}{R^3_\Sigma},
\label{ppar2}
\end{equation}
with $a_1\equiv\frac{h^\prime_\Sigma r_\Sigma}{h_\Sigma}$.

On the other hand, from (\ref{j3}), using (\ref{pa3bt}) we obtain

\begin{eqnarray}
&&2\ddot R_\Sigma R_\Sigma-\dot R^2_\Sigma-3\epsilon \alpha^2 R^4_\Sigma-4\alpha^2 \sqrt{\epsilon} R^3_\Sigma (a_1-1)\nonumber \\&-&\alpha^2R^2_\Sigma a_1(a_1-2)=0,
\label{pr0}
\end{eqnarray}

To specify a model we have to obtain $f(t)$ from the solution to the above equations.

In the special case $a_1=1$  (\ref{ppar1}) becomes
\begin{equation}
\dot R^2_\Sigma= \alpha^2 R^4_\Sigma \left[\epsilon-\frac{1}{R^2_\Sigma}+\frac{2M}{R^3_\Sigma}\right],
\label{ppar3}
\end{equation}
which has exactly the same form  as (\ref{mod17}) and therefore  admits the same  kind of solutions, and (\ref{pr0}) reads
\begin{eqnarray}
2\ddot R_\Sigma R_\Sigma-\dot R^2_\Sigma-3\epsilon \alpha^2 R^4_\Sigma+\alpha^2R^2_\Sigma =0,
\label{pr0b}
\end{eqnarray}
a first integral of which, as it can be easily shown, is (\ref{ppar3}), therefore we only need to satisfy (\ref{ppar3}).

In order to determine the functions $g(r), h(r)$ we shall assume the vanishing complexity factor condition $Y_{TF}=0$.
 
 Using (\ref{pa3bt}) in (\ref{itfm}) the condition $Y_{TF}=0$ reads

 \begin{equation}
 \frac{g^{\prime \prime}}{g^\prime}-\frac{1}{r}+\frac{2h^\prime}{h}=0,
 \label{ppar4}
 \end{equation}

or
 \begin{equation}
 \frac{v^\prime}{v}-\frac{1}{r}+\frac{2h^\prime}{h}=0,
 \label{ppar7}
 \end{equation}
with $v\equiv g^\prime$, whose formal solution is

\begin{equation}
v=\frac{c_4 r}{h^2},
 \label{ppar8}
 \end{equation}
 producing
 \begin{equation}
g=c_4\int\frac{ r}{h^2} dr+c_5,
 \label{ppar9}
 \end{equation}
where $c_4, c_5$ are arbitrary constants.

If we choose 
\begin{equation}
h(r)=c_6r,
 \label{ppar9b}
 \end{equation}
implying $a_1=1$, then we obtain from (\ref{ppar9})
\begin{equation}
g=c_7\ln r+c_5,
 \label{ppar9c}
 \end{equation}
where $c_6, c_7$ are constants.

Thus, let us consider the following model. The  time dependence  described by $f(t)$ is obtained from the solution to (\ref{ppar3}) given by

\begin{equation}\label{caso1b}
    R^{(VII)}_\Sigma=\frac{6 M\tanh ^2\left[\frac{\alpha}{2}(t-t_0)\right]}{3-\tanh ^2\left[\frac{\alpha}{2}(t-t_0)\right]},
  \end{equation}
with $\epsilon=\frac{1}{27M^2}$, and the radial dependence of the model is given by the functions $g(r), h(r)$ given by (\ref{ppar9b}) (\ref{ppar9c}).

The  physical variables corresponding to this model read

\begin{eqnarray}\label{mupal1}
&&8\pi \mu=\frac{3 \coth^2[\frac{\alpha}{2}(t-t_0)]}{4 M^2 \sinh^{4}[\frac{\alpha}{2}(t-t_0)]}+\frac{(3 \coth^2[\frac{\alpha}{2}(t-t_0)]-1)^2}{36M^2} \\&-&\frac{1}{9M^2} + \ln\left(\frac{r}{r_\Sigma}\right)\left[\frac{3 \coth^2[\frac{\alpha}{2}(t-t_0)]-1}{9\sqrt{3}M^2}+\frac{1}{27M^2}\ln\left(\frac{r}{r_\Sigma}\right)\right] \nonumber,
\end{eqnarray}

\begin{eqnarray}
8\pi P_r&=& \ln\left(\frac{r}{r_\Sigma}\right)\left[\frac{(3 \coth^2[\frac{\alpha}{2}(t-t_0)]-1)(3\coth^2[\frac{\alpha}{2}(t-t_0)]-5)}{18\sqrt{3}M^2}- \right.\nonumber  \\ & &\left. \frac{1}{27M^2} \ln\left(\frac{r}{r_\Sigma}\right)\right],
\end{eqnarray}

\begin{eqnarray}
8\pi P_\bot=\frac{(3 \coth^2[\frac{\alpha}{2}(t-t_0)]+1)}{12 M^2 \sinh^{2}[\frac{\alpha}{2}(t-t_0)]}+\frac{1}{9M^2} \nonumber \\+ \ln\left(\frac{r}{r_\Sigma}\right)\frac{(3 \coth^2[\frac{\alpha}{2}(t-t_0)]-1) }{6\sqrt{3}M^2\sinh^{2}[\frac{\alpha}{2}(t-t_0)},
\label{ptpal1b}
\end{eqnarray}
where the following relationships between the constants has been used $\alpha^2\equiv c_6^2$, $\epsilon\equiv\frac{1}{27M^2}\equiv c_7^2 c_6^2$

In the limit $t\rightarrow\infty$ the above model tends to a static fluid distribution described by 
\begin{eqnarray}
8\pi \mu=\ln\left(\frac{r}{r_\Sigma}\right)\left[\frac{2}{9\sqrt{3}M^2}+\frac{1}{27M^2}\ln\left(\frac{r}{r_\Sigma}\right)\right],
\label{mupal1s}
\end{eqnarray}

\begin{equation}
8\pi P_r=-\ln\left(\frac{r}{r_\Sigma}\right)\left[\frac{2}{9\sqrt{3}M^2}+\frac{1}{27M^2}\ln\left(\frac{r}{r_\Sigma}\right)\right],
\label{prpal1s}
\end{equation}

\begin{eqnarray}
8\pi P_\bot=\frac{1}{9M^2} ,
\label{ptpal1s}
\end{eqnarray}
satisfying the equation of state $P_r=-\mu$.

Another case which allows integration in terms of elementary function may be  obtained from the conditions  $\epsilon=0$ and $a_1=1/2$.  Then (\ref{ppar1})  reads

\begin{equation}
\dot R^2_\Sigma= \alpha^2 R^4_\Sigma \left[\frac{2M}{R^3_\Sigma}-\frac{3}{4R^2_\Sigma}\right].
\label{pal1}
\end{equation}

The above equation may be easily integrated, producing
\begin{equation}
R^{(VIII)}_\Sigma= \frac{4M}{3} \left(1 + \sin \tilde t\right),
\label{ppar15}
\end{equation}
with $\tilde t\equiv  \frac{\sqrt{3}\alpha}{2}(t-t_0)$.

Next, in order to specify further the model, we shall impose the vanishing complexity factor condition. In this case  ($a_1=1/2$), the general solution to (\ref{ppar7}) reads \begin{equation}
h=c_1\sqrt{r}, \qquad  g=c_2r+c_3,
 \label{ppar13}
 \end{equation}
however since $\epsilon=0$ the constant $c_2$ must vanish.

The physical variables for this model read

\begin{eqnarray}
8\pi \mu=\frac{27}{64M^2(1+ \sin\tilde t)^2}\left[\frac{3 \cos^2 \tilde t}{(1+ \sin \tilde t)^2}+\frac{r_\Sigma}{r}\right],
\label{mue0}
\end{eqnarray}

\begin{eqnarray}
8\pi P_r=\frac{27}{64M^2(1+ \sin\tilde t)^2}\left(1-\frac{r_\Sigma}{r}\right),
\label{pre0}
\end{eqnarray}

\begin{eqnarray}
8\pi P_\bot=\frac{27}{64M^2 (1+ \sin\tilde t)^2}.
\label{pte0}
\end{eqnarray}

This solution represents  fluid distribution oscillating between $R_\Sigma=0$  and  $R_\Sigma=\frac{8M}{3}$. It is worth noticing that the energy density is always positive, whereas the radial pressure is not.

 Finally, we shall present two solutions describing  a ``ghost'' compact object, of the kind already discussed in the previous section.

Thus assuming $M=0$, equation (\ref{ppar1}) becomes
\begin{equation}
\dot R^2_\Sigma= \alpha^2 R^4_\Sigma \left[\epsilon-\frac{2\sqrt{\epsilon}(1-a_1)}{R_\Sigma}-\frac{a_1(2-a_1)}{R^2_\Sigma}\right].
\label{ppar16}
\end{equation}

Solutions to the above equation in terms of elementary functions may be obtained by assuming $a_1=1$, in which case the two possible solutions to (\ref{ppar16}) are

\begin{equation}
R^{(IX)}_\Sigma=\frac{1}{\sqrt{\epsilon} \cos[\alpha(t-t_0)]},
\label{ppar17}
\end{equation}
and
\begin{equation}
R^{(X)}_\Sigma=\frac{1}{\sqrt{\epsilon} \sin[\alpha(t-t_0)]}.
\label{ppar18}
\end{equation}

Imposing further the vanishing complexity factor condition, then functions $h(r), g(r)$ are given by (\ref{ppar9b}) and (\ref{ppar9c}).
The physical variables corresponding to (\ref{ppar17}) and (\ref{ppar18}) read respectively

\begin{eqnarray}
8\pi \mu&=&-2 \epsilon \cos^2[\alpha (t-t_0)]+\nonumber \\&+&\epsilon \ln\frac{r}{r_\Sigma}\left[2 \cos[\alpha(t-t_0)]+\ln\frac{r}{r_\Sigma}\right],
\label{mum0}
\end{eqnarray}

\begin{eqnarray}
8\pi P_r=-\epsilon \ln\frac{r}{r_\Sigma}\left[4 \cos[\alpha(t-t_0)]+\ln\frac{r}{r_\Sigma}\right],
\label{pprm0}
\end{eqnarray}

\begin{equation}
8\pi P_\bot=\epsilon \cos^2[\alpha (t-t_0)]- 2\epsilon \ln\frac{r}{r_\Sigma}\cos[\alpha(t-t_0)],
\label{pptm0}
\end{equation}
and 

\begin{eqnarray}
8\pi \mu&=&-2 \epsilon \sin^2[\alpha (t-t_0)]+\nonumber \\&+&\epsilon \ln\frac{r}{r_\Sigma}\left[2 \sin[\alpha(t-t_0)]+\ln\frac{r}{r_\Sigma}\right],
\label{mum0}
\end{eqnarray}

\begin{eqnarray}
8\pi P_r=-\epsilon \ln\frac{r}{r_\Sigma}\left[4 \sin[\alpha(t-t_0)]+\ln\frac{r}{r_\Sigma}\right],
\label{pprm0}
\end{eqnarray}

\begin{equation}
8\pi P_\bot=\epsilon \sin^2[\alpha (t-t_0)]- 2\epsilon \ln\frac{r}{r_\Sigma}\sin[\alpha(t-t_0)].
\label{pptm0}
\end{equation}

\subsection{$\chi^\alpha \vert \vert V^\alpha; q\neq 0$}
Finally, we shall consider the case where the CKV is parallel to the four--velocity, and the system is dissipative.
As result of the admittance of the CKV the metric functions read as (\ref{pa1}). Then, feeding this back into (\ref{13})  produces

\begin{equation}
\frac{\dot B^\prime}{B}-2\frac{\dot B}{B}\frac{B^\prime}{B}-\frac{\dot B}{B}\frac{h^\prime}{h}=4\pi q A B,
\label{chipd1}
\end{equation}
which may be formally integrated, to obtain

\begin{equation}
B=\frac{1}{h(r)\left[f(t)+g(r)-\int \int 4\pi q B dr dt \right]},
\label{chipd2}
\end{equation}
implying
\begin{equation}
A=\frac{1}{\left[f(t)+g(r)-\int \int 4\pi q B dr dt \right]},
\label{chipd3}
\end{equation}
and 
\begin{equation}
R=\frac{r}{h(r)\left[f(t)+g(r)-\int \int 4\pi q B dr dt \right]},
\label{chipd4}
\end{equation}
where $f(t)$ and $g(r)$ are arbitrary  functions of their  arguments.

To specify a model we shall impose the vanishing complexity factor condition. 
Thus, using (\ref{chipd2})--(\ref{chipd4}) in (\ref{itfm}) the condition $Y_{TF}=0$ reads
\begin{equation}
g^\prime-\int{4\pi q Bdt}=\frac{\gamma(t)r}{h^2(r)},
\label{chipd5}
\end{equation}
a formal integration of which produces 
\begin{equation} 
g-\int{\int{4\pi q Bdt dr}}=\gamma(t)\int{\frac{rdr }{h^2(r)}},
\label{chipd7}
\end{equation}
where $\gamma(t)$ is an arbitrary function.

Also, taking the $t$-derivative of (\ref{chipd5}) we obtain
\begin{equation}
4\pi q B=-\frac{\dot\gamma(t)r}{h^2(r)}.
\label{chipd6}
\end{equation}

Using the above expressions we may write the metric functions (\ref{chipd2})--(\ref{chipd4}) as

\begin{equation}
B=\frac{1}{h(r)\left[f(t)+\gamma(t)\int{\frac{rdr }{h^2(r)}}\right]},
\label{chipd2b}
\end{equation}
implying
\begin{equation}
A=\frac{1}{\left[f(t)+\gamma(t)\int{\frac{rdr }{h^2(r)}} \right]},
\label{chipd3b}
\end{equation}
and 
\begin{equation}
R=\frac{r}{h(r)\left[f(t)+\gamma(t)\int{\frac{rdr }{h^2(r)}}\right]}.
\label{chipd4b}
\end{equation}

Further restrictions on functions $f(t), \gamma(t)$ will be obtained from the junction condition $(P_r=q)_\Sigma$.

Indeed, using (\ref{chipd6})--(\ref{chipd4b}) and (\ref{14}), the condition  $(P_r=q)_\Sigma$, reads
\begin{eqnarray} 
&&h^2_\Sigma X_\Sigma\left(\frac{1}{r_\Sigma}-\frac{h^\prime_\Sigma}{h_\Sigma}-\frac{X^\prime_\Sigma}{X_\Sigma}\right) \left(\frac{1}{r_\Sigma}-\frac{h^\prime_\Sigma}{h_\Sigma}-3\frac{X^\prime_\Sigma}{X_\Sigma}\right)\nonumber \\&-&X_\Sigma\left(-2\frac{\ddot X_\Sigma}{X_\Sigma}+3\frac{\dot X^2_\Sigma}{X^2_\Sigma}+\alpha^2 \right)
=-\frac{2 \dot\gamma}{\alpha}
\label{pqs1},
\end{eqnarray}
where 
\begin{equation}
X \equiv f(t)+\gamma(t)\int{\frac{rdr }{h^2(r)}}, \qquad \alpha \equiv\frac{h_\Sigma}{r_\Sigma}.
\label{pqs2}
\end{equation}

In order to solve the above equation we shall assume
\begin{eqnarray}
-2\frac{\ddot X_\Sigma}{X_\Sigma}+3\frac{\dot X^2_\Sigma}{X^2_\Sigma}+\alpha^2 =\frac{2 \dot\gamma}{X_\Sigma\alpha},
\label{psq3}
\end{eqnarray}
and
\begin{equation}
1-a_1-\frac{r_\Sigma X^\prime_\Sigma}{X_\Sigma}=0,
\label{psq4}
\end{equation}
where $a_1\equiv \frac{h^\prime_\Sigma r_\Sigma}{h_\Sigma}$.

From (\ref{psq4}) it follows at once that 

\begin{equation}
1-a_1=\frac{\gamma(t)}{\alpha^2 X_\Sigma},
\label{psq5}
\end{equation}
producing
\begin{equation}
\frac{\gamma(t)\dot X_\Sigma}{ X^2_\Sigma}=\frac{\dot \gamma(t)}{X_\Sigma}.
\label{psq6}
\end{equation}

Using (\ref{psq5}) and (\ref{psq6}) in (\ref{psq3}), this last equation becomes
\begin{eqnarray}
2\frac{\ddot X_\Sigma}{X_\Sigma}-3\frac{\dot X^2_\Sigma}{X^2_\Sigma}+\frac{2\alpha \dot X_\Sigma (1-a_1)}{X_\Sigma}-\alpha^2 =0.
\label{psq7}
\end{eqnarray}

\noindent In order to integrate (\ref{psq7}), let us introduce the  variable $y=\frac{\dot{X}_\Sigma}{X_\Sigma}$, in terms of which (\ref{psq7}) reads

\begin{equation}\label{ecy}
  \dot{y}-\frac{1}{2}y^2+\alpha (1-a_1) y-\frac{\alpha ^2}{2}=0.
\end{equation}

\noindent This is a  Ricatti equation, a particular solution of which is 
\begin{equation}\label{y0}
  y_0=\alpha(1-a_1)\pm \alpha \sqrt{a^2_1-2a_1}.
\end{equation}

\noindent Then, in order to find the general solution to (\ref{ecy}) let us introduce the variable  $z=y-y_0$, producing

\begin{equation}\label{ecz}
\dot{z}-\frac{1}{2}z^2+\left[\alpha (1-a_1)-y_0\right]z=0,
\end{equation}
whose solution reads

\begin{equation}\label{solz}
z=\frac{2 \delta}{1+be^{\delta t}},
\end{equation}
where $b$ is an arbitrary constant of integration and $\delta \equiv \alpha(1-a_1)-y_0$.

With this result, we can easily find $X_\Sigma$, whose expression reads
\begin{equation}\label{solx}
X_\Sigma=\frac{c e^{(2\delta+y_0)t}}{(1+be^{\delta t})^2},
\end{equation}
where $c$ is a constant of integration.

Using (\ref{solx}) in (\ref{psq5}) we obtain the explicit form of $\gamma(t)$, and using this expression and (\ref{solx}) in (\ref{pqs2}) we obtain the explicit from of $f(t)$. Thus, the model is completely determined up to a single function of $r$ ($h(r)$).

In terms of $X(t,r)$ and $h(r)$, the physical variables read
\begin{eqnarray}
8\pi \mu=& &3\dot X^2-h^2X^2\left[-\frac{2h^{\prime \prime}}{h}+\frac{3(h^\prime)^2}{h^2}-\frac{2 X^{\prime \prime}}{X}+\frac{3(X^\prime)^2}{X^2}\nonumber \right.\\& &\left.+\frac{2 h^\prime X^\prime}{hX}  - \frac{4h^\prime}{rh}-\frac{4X^\prime}{rX}\right],
\label{mues3}
\end{eqnarray}
\begin{eqnarray}
8\pi P_r=& &2\ddot X X-3\dot X^2+h^2X^2\left[\frac{(h^\prime)^2}{h^2}+\frac{3(X^\prime)^2}{X^2}+\frac{4 h^\prime X^\prime}{hX} \nonumber \right.\\ & & \left. - \frac{2h^\prime}{rh}-\frac{4X^\prime}{rX}\right],
\label{pres3}
\end{eqnarray}

\begin{eqnarray}
8\pi P_\bot=& &2\ddot X X-3\dot X^2+h^2X^2\left[-\frac{h^{\prime \prime}}{h}+\frac{(h^\prime)^2}{h^2}-\frac{2 X^{\prime \prime}}{X}\nonumber \right. \\& &\left.+\frac{3(X^\prime)^2}{X^2}  - \frac{h^\prime}{rh}-\frac{2X^\prime}{rX}\right],
\label{ptes3}
\end{eqnarray}

\begin{eqnarray}
8\pi q= -2 h X \dot X^{\prime}.
\label{qes3}
\end{eqnarray}

In order to obtain a specific model we shall assume $a_1=2$, which implies $y_0=-\alpha$ and $\delta=0$, then feeding back these values in (\ref{solx}), the expression for $X_\Sigma$ becomes
\begin{equation}\label{solxsp1}
X_\Sigma=\tilde c e^{-\alpha t},
\end{equation}
with $\tilde c \equiv \frac{c}{(1+b)^2}$.

Next, we shall assume for $h(r)$ the form
\begin{equation}
h(r)=c_2r^2,
\label{ces2}
\end{equation}
where $c_2$ is a constant with dimensions  $1/[length]^2$, producing
\begin{equation}
\alpha=c_2r_\Sigma.
\label{ces3}
\end{equation}

Using (\ref{ces2}) and (\ref{ces3}) we obtain for the radial dependence of $X$

\begin{equation}
\int{\frac{r dr}{h(r)^2}}=-\frac{r_\Sigma^2}{2 \alpha^2 r^2}.
\label{ces4}
\end{equation}

From (\ref{psq5}) we obtain at once for $\gamma(t)$

\begin{equation}
\gamma(t)=-\alpha^2 \tilde c e^{-\alpha t},
\label{ces5}
\end{equation}
and from (\ref{pqs2}) and (\ref{ces4}) the expression for $f(t)$ reads
\begin{equation}
f(t)=\frac{\tilde c e^{-\alpha t}}{2}.
\label{ces5}
\end{equation}

Finally the expression for $X(t,r)$ reads

\begin{equation}
X^{(XI)}(t,r)=\frac{\tilde c e^{-\alpha t}}{2}\left[1+\left(\frac{r_\Sigma}{r}\right)^2\right].
\label{ces6}
\end{equation}

Thus, the physical variables for this model $XI$ (including the total mass and the temperature) read
\begin{equation}
8\pi \mu=\frac{3\tilde c^2\alpha^2e^{-2\alpha t}}{4r^4}\left(5r^4+2r^2r^2_\Sigma+r^4_\Sigma\right),
\label{mdp}
\end{equation}

\begin{equation}
8\pi P_r=8\pi P_\bot=\frac{\tilde c^2\alpha^2e^{-2\alpha t}}{4r^4}\left(2r^2r^2_\Sigma-9r^4-r^4_\Sigma\right),
\label{pdp}
\end{equation}

\begin{equation}
8\pi q=-\tilde c^2\alpha^2e^{-2\alpha t}\left(r^2+r^2_\Sigma\right)\frac{r_\Sigma}{r^3},
\label{qdp}
\end{equation}

\begin{equation}
m_\Sigma=\frac{e^{\alpha t}}{\tilde c\alpha},
\label{mdp}
\end{equation}

\begin{eqnarray}
T(t,r)=\frac{\tilde c e^{-\alpha t}(r^2+r^2_\Sigma)}{2r^2}\left[\frac{\tau r^2_\Sigma\tilde c\alpha^2e^{-\alpha t}}{4\pi \kappa r^2}\right.  \nonumber \\ \left.+\frac{\alpha}{4\pi \kappa} \ln\left(\frac{r^2}{r^2+r^2_\Sigma}\right)+T_0(t)\right],
\label{tempdp}
\end{eqnarray}
this last expression was obtained using the truncated transport equation (\ref{5trun}). 

It is worth noticing that this model is intrinsically isotropic in pressure, the energy density is positive and larger than the pressure, and the matching condition $q\stackrel{\Sigma}{=}P_r$ is obviously satisfied. However the physical variables are singular at the center.
\section{Discussion}
We have seen so far that the admittance of CKV leads to a wealth of  solutions to the Einstein equations for a general spherically symmetric fluid distributions, which could be applied to a variety of astrophysical problems, or serve as testbeds for discussions about theoretical issues such as wormholes and white holes. 

In order to find solutions expressed in terms of elementary functions  we have imposed further constrains on the fluid distribution. Some of which are endowed with a distinct physical meaning (e.g. the vanishing complexity factor, or the quasi--homologous condition), while others have been imposed just to produce models described by elementary functions. 

We started by considering non--dissipative fluids admitting a CKV orthogonal to the four--velocity. In this case the assumed symmetry reduces the metric variables to three functions (two functions  of $t$ and one function of $r$).  Then, the matching conditions reduce to a single differential equation (\ref{mod17}) whose solution provides one of the three  functions describing the metric. In order to obtain a solution expressed in terms of elementary functions we have  assumed  specific values of the parameters entering into the equation. 

The first choice ($\omega=\frac{1}{27M^2}$) leads to two expressions for the areal radius of the boundary  ((\ref{caso1}) and (\ref{caso1ba})). The first one describes a fluid distribution whose boundary areal radius  expands from $0$ to $3M$, while the second one describes a contraction  of the boundary areal radius from infinity to $3M$. To find the remaining two functions to determine the metric we have  assumed the quasi--homologous condition and the vanishing complexity factor condition. In this way we are lead to our models $I$ and $II$, both of which have positive energy densities and the physical variables are singular free, except the model $I$ for $t=t_0$. 

As $t\rightarrow\infty$ both solutions tend to the same static solution (\ref{sl}) satisfying  a   Chaplygin--type equation  of state $\mu=-P_r$. The way of reaching this static limit deserves some comments. Usually the hydrostatic equilibrium is reached when the ``gravitational force term'' (the first term on the right of (\ref{3m})) cancels the ``hydrodynamic force term'' (the second term on the right of  (\ref{3m})). However here the situation is different, the equilibrium is reached because as $t\rightarrow\infty$ both terms tend to zero. 

In spite of the good behavior of  these two models, it should be mentioned that regularity conditions are not satisfied by the resulting function $R$ on the center of the distribution. Accordingly for the modeling of any specific scenario, the central region should be excluded.

Next, we have considered the case $\omega=0$, which together  with the vanishing complexity factor condition produces the model $III$. In this model the boundary areal radius oscillates between  $0$ and $2M$. The energy density  and the tangential pressure of this model are positive and homogeneous, while the radial pressure vanishes identically. As in the previous two models this solution does not satisfy the regularity condition at the center. 

As an additional  example of  analytical solution we have considered  the case $M=0$. The two models for this kind of solution are the models $IV$ and $V$. They represent a kind of ``ghost'' stars, formed  by a fluid distribution not producing gravitational effects outside the boundary surface. They present  pathologies, both physical and topological,  and therefore their physical applications are dubious. However since this kind of distributions have been considered in the past (see for example \cite{zel}) we present them here.

Next we have considered the subcase where the  CKV is orthogonal to the four velocity and the fluid is dissipative. For this case we have found a model satisfying the vanishing complexity factor and the quasi--homologous condition, which together with the fulfillment of the matching conditions determine all the metric functions. This model (model $VI$) is described by expressions (\ref{metricsf1})--(\ref{ptidscosf}), and the expression (\ref{tem}) for the temperature, which has been calculated using the truncated version of the transport equation. It contains contribution from the transient regime (proportional  to $\tau$) as well as from the stationary regime. As previous models, this solution does not satisfy the regularity conditions at the center.

The other family of solutions corresponds to the case when the  CKV is parallel to the four--velocity. In the non--dissipative case, as consequence of this  symmetry,  the metric functions 
are determined  up to three functions (two functions of $r$ and one function of $t$). Besides, the fluid is necessarily shear--free,  a result which was already known \cite{t1,t3}.
The function of $t$ is obtained from the fulfillment of the matching conditions (Eqs. (\ref{ppar1}), (\ref{pr0})). These equations have been integrated for different values of the parameters entering into them. Thus, for $a_1=1$ and $\epsilon=\frac{1}{27M^2}$, together with the vanishing complexity factor condition and $h(r)=c_6r$, we have found model $VII$. The boundary areal radius of this model expands from zero to $3M$, and the physical variables are given by (\ref{mupal1})--(\ref{ptpal1b}).  In the limit $t\rightarrow\infty$ the model tends to a static sphere whose equation of state is $P_r=-\mu$. The energy density is positive, and presents  a singularity only at $t=t_0$, however regularity conditions are not satisfied at the center.

The integration of the matching conditions for $\epsilon=0$ and $a_1=\frac{1}{2}$ together with the vanishing complexity factor, produce the model $VIII$. The boundary areal radius of this model oscillates between zero and $\frac{8M}{3}$. The energy density is positive and larger than the radial pressure, but the fluid distribution is singular at $r=0$.

For $M=0$ and $a_1=1$ we obtain models $IX$ and $X$ they describe the kind of ``ghost stars'' mentioned before. However they are plagued with, both, physical and topological pathologies which renders them unviable for physical modeling. We include them just for sake of completeness.

Finally, we have considered  the dissipative case for the CKV parallel to the four--velocity. The metric variables for this case  take the form (\ref{chipd2})--(\ref{chipd4}), which after imposing the vanishing complexity factor condition become (\ref{chipd2b})--(\ref{chipd4b}). Thus the metric is determined up to three functions (two functions of $t$ and one function of $r$). The two functions of $t$ will be obtained from the integration of the matching conditions, while the function of $r$ is assumed as (\ref{ces2}). The model is further specified with the choice $a_1=2$. This produce the model $XI$.

 As it follows from (\ref{ces6}) the boundary areal radius of the model tends to infinity as $t\rightarrow\infty$, while in the same limit the total mass $m_\Sigma$ tends to infinity, whereas  both $q$ and $\mu$ tend to zero. The explanation for this strange result comes about from the fact that $R_\Sigma$ grows exponentially with $t$, overcompensating the decreasing of $\mu$ and $q$ in (\ref{27intcopy}). It is also worth noticing the negative sign of $q$, implying an inward heat flux driving the expansion of the fluid distribution.

Overall, we believe that the eleven models exhibited (or at least some of them)  could be useful to describe some stages of some regions  of  self--gravitating fluid, in the  evolution of compact objects. Each specific scenario imposing specific values on the relevant parameters. It should be reminded that in any realistic collapsing scenario we do not expect the same equation of state to be valid all along the evolution and for the whole fluid configuration.

Before concluding, some  general comments are in order.
\begin{enumerate}
\item The analytical integration of the equations derived from the matching conditions have been carried out by imposing specific values on the parameters entering into those equations, also the models have been specified by using some conditions such as the quasi--homologous condition. Of course the number of available options is huge. Among them we would like to mention the prescription of the total luminosity measured by  an observer at rest at infinity  (\ref{lum}). Let us recall that this is one of the few observables in the process of stellar evolution. Equivalently one could propose a specific evolution of the total mass with time.
\item In some cases, when the topological pathologies are not ``severe'', the time interval of viability of the solution may be restricted by the condition that $U\leq 1$ (e.g. for solutions $I$ and $II$). In other cases however, due to topological defects, the interpretation of $U$ as a velocity becomes dubious and  therefore it is not clear that $U$ should satisfy the above mentioned condition.
\item Model $XI$ is dissipative and intrinsically isotropic in pressure. However as shown in \cite{ps} dissipation produces pressure anisotropy, unless a highly unlikely cancellation of the four terms on the right of equation (28) in \cite{ps} occurs. This happens in model $XI$, which renders  this solution a very remarkable one.
\item For reasons explained in the Introduction we have focused on the obtention of analytical solutions expressed through elementary functions. However it should be clear that for specific astrophysical  scenarios, a numerical approach for solving the matching conditions, could be more appropriate.
\end{enumerate}
\begin{acknowledgments} 

This  work  was partially supported by the Spanish  Ministerio de Ciencia e
Innovaci\'on under Research Projects No.  FIS2015-65140-P (MINECO/FEDER). ADP acknowledges hospitality from the Physics Department of  the Universitat de les Illes Balears.

\end{acknowledgments}
\appendix
\section{Einstein equations}
 Einstein's field equations for the interior spacetime (\ref{1}) are given by
\begin{equation}
G_{\alpha\beta}=8\pi T_{\alpha\beta},
\label{2}
\end{equation}
and its non zero components
read
\begin{widetext}
\begin{eqnarray}
8\pi T_{00}=8\pi  \mu A^2
=\left(2\frac{\dot{B}}{B}+\frac{\dot{R}}{R}\right)\frac{\dot{R}}{R}
-\left(\frac{A}{B}\right)^2\left[2\frac{R^{\prime\prime}}{R}+\left(\frac{R^{\prime}}{R}\right)^2
-2\frac{B^{\prime}}{B}\frac{R^{\prime}}{R}-\left(\frac{B}{R}\right)^2\right],
\label{12} \\
8\pi T_{01}=-8\pi qAB
=-2\left(\frac{{\dot R}^{\prime}}{R}
-\frac{\dot B}{B}\frac{R^{\prime}}{R}-\frac{\dot
R}{R}\frac{A^{\prime}}{A}\right),
\label{13} \\
8\pi T_{11}=8\pi P_r B^2
=-\left(\frac{B}{A}\right)^2\left[2\frac{\ddot{R}}{R}-\left(2\frac{\dot A}{A}-\frac{\dot{R}}{R}\right)
\frac{\dot R}{R}\right]
+\left(2\frac{A^{\prime}}{A}+\frac{R^{\prime}}{R}\right)\frac{R^{\prime}}{R}-\left(\frac{B}{R}\right)^2,
\label{14} \\
8\pi T_{22}=\frac{8\pi}{\sin^2\theta}T_{33}=8\pi P_{\perp}R^2
=-\left(\frac{R}{A}\right)^2\left[\frac{\ddot{B}}{B}+\frac{\ddot{R}}{R}
-\frac{\dot{A}}{A}\left(\frac{\dot{B}}{B}+\frac{\dot{R}}{R}\right)
+\frac{\dot{B}}{B}\frac{\dot{R}}{R}\right]\nonumber \\
+\left(\frac{R}{B}\right)^2\left[\frac{A^{\prime\prime}}{A}
+\frac{R^{\prime\prime}}{R}-\frac{A^{\prime}}{A}\frac{B^{\prime}}{B}
+\left(\frac{A^{\prime}}{A}-\frac{B^{\prime}}{B}\right)\frac{R^{\prime}}{R}\right].\label{15}
\end{eqnarray}
\end{widetext}
The component (\ref{13}) can be rewritten with (\ref{5c1}) and
(\ref{5b}) as
\begin{equation}
4\pi qB=\frac{1}{3}(\Theta-\sigma)^{\prime}
-\sigma\frac{R^{\prime}}{R}.\label{17a}
\end{equation}
\section{Dynamical equations}

The non trivial components of the Bianchi identities, $T^{\alpha\beta}_{;\beta}=0$, from (\ref{2}) yield
\begin{widetext}
\begin{eqnarray}
T^{\alpha\beta}_{;\beta}V_{\alpha}=-\frac{1}{A}\left[\dot { \mu}+
\left( \mu+ P_r\right)\frac{\dot B}{B}
+2\left( \mu+P_{\perp}\right)\frac{\dot R}{R}\right]
-\frac{1}{B}\left[ q^{\prime}+2 q\frac{(AR)^{\prime}}{AR}\right]=0, \label{j4}\\
T^{\alpha\beta}_{;\beta}K_{\alpha}=\frac{1}{A}\left[\dot { q}
+2 q\left(\frac{\dot B}{B}+\frac{\dot R}{R}\right)\right]
+\frac{1}{B}\left[ P_r^{\prime}
+\left(\mu+ P_r \right)\frac{A^{\prime}}{A}
+2( P_r-P_{\perp})\frac{R^{\prime}}{R}\right]=0, \label{j5}
\end{eqnarray}
\end{widetext}
or, by using (\ref{5c}), (\ref{5c1}), (\ref{16}), (\ref{23a}) and (\ref{20x}), they become,
\begin{widetext}
\begin{eqnarray}
D_T \mu+\frac{1}{3}\left(3 \mu+ P_r+2P_{\perp} \right)\Theta
+\frac{2}{3}( P_r-P_{\perp})\sigma+ED_R q
+2 q\left(a+\frac{E}{R}\right)=0, \label{j6}
\end{eqnarray}
\begin{eqnarray}
D_T q+\frac{2}{3} q(2\Theta+\sigma)
+ED_R  P_r
+\left( \mu+ P_r \right)a+2(P_r-P_{\perp})\frac{E}{R}=0.
\label{j7}
\end{eqnarray}
\end{widetext}

This last equation may be further transformed as follows, the acceleration $D_TU$ of an infalling particle can
be obtained by using (\ref{5c}), (\ref{14}), (\ref{17masa})  and (\ref{20x}),
producing
\begin{equation}
D_TU=-\frac{m}{R^2}-4\pi  P_r R
+Ea, \label{28}
\end{equation}
and then, substituting $a$ from (\ref{28}) into
(\ref{j7}), we obtain
\begin{widetext}
\begin{eqnarray}
\left( \mu+ P_r\right)D_TU
=-\left(\mu+ P_r \right)
\left[\frac{m}{R^2}
+4\pi  P_r R\right]
-E^2\left[D_R  P_r
+2(P_r-P_{\perp})\frac{1}{R}\right]
-E\left[D_T q+2 q\left(2\frac{U}{R}+\sigma\right)\right].
\label{3m}
\end{eqnarray}
\end{widetext}

\end{document}